\documentclass[
superscriptaddress,
aps,
pra,
twocolumn,
longbibliography
]{revtex4-1}

\usepackage{bm}
\usepackage{ulem}
\usepackage{epsfig}
\usepackage{graphicx}
\usepackage{amssymb,amsmath,amsbsy,amsgen,amsfonts}    
\usepackage{dcolumn}
\usepackage{amsthm}
\usepackage{mathrsfs}
\usepackage{latexsym}
\usepackage{array}
\usepackage{color}
\usepackage{amstext}
\allowdisplaybreaks[1]
\usepackage{txfonts}
\usepackage{pifont}
\usepackage{braket}
\usepackage{epstopdf}
\usepackage{footnote}
\usepackage{hyperref}

\usepackage{setspace} 

\DeclareSymbolFont{symbols}{OMS}{cmsy}{m}{n}

\begin{document}
\title{Superconducting nanowire single-photon spectrometer exploiting cascaded photonic crystal cavities}
\author{Youngsun Yun}%
\affiliation{Institute of Theoretical Solid State Physics, Karlsruhe Institute of Technology, 76131 Karlsruhe, Germany}
\affiliation{Advanced Photonics Research Institute, Gwangju Institute of Science and Technology, Gwangju 61005, Republic of Korea}

\author{Andreas Vetter}
\affiliation{Institute of Nanotechnology, Karlsruhe Institute of Technology, 76021 Karlsruhe, Germany}
\affiliation{SUSS MicroOptics SA, Rouges-Terres 61, CH-2068 Hauterive, Switzerland}

\author{Robin Stegmueller}
\affiliation{University of M\"unster, Institute of Physics, Wilhelm-Klemm-Str. 11, 48149 M\"unster, Germany}
\affiliation{University of M\"unster, CeNTech Center for Nanotechnology, Heisenbergstr. 11, 48149 M\"unster, Germany}

\author{Simone Ferrari}
\affiliation{University of M\"unster, Institute of Physics, Wilhelm-Klemm-Str. 11, 48149 M\"unster, Germany}
\affiliation{University of M\"unster, CeNTech Center for Nanotechnology, Heisenbergstr. 11, 48149 M\"unster, Germany}

\author{\\Wolfram H. P. Pernice}
\affiliation{University of M\"unster, Institute of Physics, Wilhelm-Klemm-Str. 11, 48149 M\"unster, Germany}
\affiliation{University of M\"unster, CeNTech Center for Nanotechnology, Heisenbergstr. 11, 48149 M\"unster, Germany}

\author{Carsten Rockstuhl}
\email{carsten.rockstuhl@kit.edu}
\affiliation{Institute of Theoretical Solid State Physics, Karlsruhe Institute of Technology, 76131 Karlsruhe, Germany}
\affiliation{Institute of Nanotechnology, Karlsruhe Institute of Technology, 76021 Karlsruhe, Germany}

\author{Changhyoup Lee}
\email{changhyoup.lee@gmail.com}
\affiliation{Institute of Theoretical Solid State Physics, Karlsruhe Institute of Technology, 76131 Karlsruhe, Germany}

\date{\today}

\begin{abstract} 
Superconducting nanowire single-photon detectors promise efficient ($\sim$100\%) and fast ($\sim$Gcps) detection of light at the single-photon level. They constitute one of the building blocks to realize integrated quantum optical circuits in a waveguide architecture. The optical response of single-photon detectors, however, is limited to measure only the presence of photons. It misses the capability to resolve the spectrum of a possible broadband illumination. In this work, we propose the optical design for a superconducting nanowire single-photon spectrometer in an integrated optical platform. We exploit a cascade of cavities with different resonance wavelengths side-coupled to a photonic crystal bus waveguide. This allows to demultiplex different wavelengths into different spatial regions, where individual superconducting nanowires that measure the presence of single photons are placed next to these cavities. We employ temporal coupled-mode theory to derive the optimal conditions to achieve a high absorption efficiency in the nanowire with fine spectral resolution. It is shown that the use of a mirror at the end of the cascaded system that terminates the photonic crystal bus waveguide increases the absorption efficiency up to unity, in principle, in the absence of loss. The expected response is demonstrated by full-wave simulations for both two-dimensional and three-dimensional structures. Absorption efficiencies of about $80\%$ are achieved both in two-dimensional structures for four cascaded cavities and in three-dimensional structures for two cascaded cavities. The achieved spectral resolution is about $1~\text{nm}$. We expect that the proposed setup, both analytically studied and numerically demonstrated in this work, offers a great impetus for future quantum nanophotonic on-chip technologies.

\end{abstract}


\maketitle

\hyphenation{he-te-ro-in-ter-face}
\hyphenation{mul-ti-pix-el}

\section{Introduction}

Single-photon detectors play an essential role in many quantum optical applications~\cite{Hadfield2009, Eisaman2011}, including quantum key distribution~\cite{Gisin2002}, linear optical quantum computing~\cite{Knill2001,Kok2007}, quantum imaging~\cite{Lugiato2002}, and quantum sensing~\cite{Pirandola2018}. A prominent single-photon detector with a great success in commercialization is a superconducting nanowire single-photon detector~(SNSPD)~\cite{Goltsman2001}. It is widely exploited because of its remarkable performance both in terms of efficiency (\text{$\sim100\%$}) and timing ($\sim$Gcps) characteristics as compared to, e.g., single-photon avalanche diodes~\cite{Eisaman2011}. Various configurations of SNSPDs have been suggested so far. Examples are the meander wires~\cite{Verevkin2002}, the meander structures implemented in optical cavities with back-reflector mirrors~\cite{Rosfjord2006}, multipixel SNSPDs~\cite{Dauler2009}, and waveguide integrated SNSPDs~\cite{Sprengers2011}. Among those, SNSPDs integrated in a photonic chip have emerged as a useful candidate geared towards building integrated quantum nanophotonic devices~\cite{Ferrari2018,Khasminskaya2016}. A reduced recovery time and dark count rate were observed with much shorter nanowires embedded in a one-dimensional~(1D) photonic crystal~(PhC) cavity~\cite{Akhlaghi2015, Vetter2016}. Dead times in the order of only 200 ps come in reach. The on-chip detection efficiency can also be further increased, in principle up to unity, by implementing SNSPDs in a two-dimensional~(2D) PhC slab~\cite{Munzberg2018}. 

The detection of single photons in SNSPDs relies on the transient breaking of the superconducting state in the superconducting nanowire, triggered by the absorption of light~\cite{Engel2015}. The response of SNSPDs indicates the presence of one or multiple photons, but misses the capability to resolve different wavelengths of light when detected~\cite{Renema2013, Engel2015}. However, access to spectral information is required to measure discrete multicolor entangled light~\cite{Ramelow2009} or spectral properties of multiphoton quantum interference~\cite{Tillmann2015}, and for being used in quantum white-light interferometry~\cite{Kaiser2018}. A standard method resolving wavelengths of light is using gratings. \text{Kahl~\textit{et al.}} have implemented an arrayed waveguide grating with different optical lengths to demultiplex distinct wavelengths of light into separated SNSPDs~\cite{Kahl2017}.
Very recently, \text{Cheng~\textit{et al.}} have reported an experimental demonstration of a broadband on-chip single-photon spectrometer using a dispersive echelle grating with a single meandered SNSPD~\cite{Cheng2019}.

The recent advances of quantum nanophotonic technology require a single-photon spectrometer to be implemented in a platform that can also accommodate other functionalities. One promising platform is provided by 2D PhC slabs, which enable on-demand manipulation of photons in various ways~\cite{OBrien2009, Englund2009}. Interaction with quantum dots has been achieved in a well-controlled manner~\cite{Vuckovic2003}, leading to the mo\-di\-fi\-ca\-tion of the spontaneous emission rate~\cite{Englund2005}, reflectivity-controllable cavities~\cite{Englund2007a}, generation and transfer of nonclassical light~\cite{Englund2007b,Vuckovic2006}, and a strong nonlinear response at the single-photon level~\cite{Javadi2015}. Engineering PhC structures has facilitated atom-atom interaction in subwavelength optical lattices~\cite{Gonzalez2015, Yu2018} and chiral light-matter interaction~\cite{Lodahl2017}. 

In this work, we propose a single-photon spectrometer using SNSPDs implemented in cascaded 2D PhC structures with different lattice constants in a slab geometry, as exemplarily illustrated in Fig.~\ref{setup}. The capability to resolve the wavelength is enabled by a wavelength-division multiplexing drop filter~\cite{Takano2005, Song2005b, Takano2006}. The latter consists of a bus waveguide coupled to an in-plane array of PhC nanocavities with distinct resonance wavelengths. The superconducting nanowires are placed next to the individual cavities and absorb the light captured in the cavities through the evanescent coupling between the cavity and the nanowire, consequently triggering detection.

For the purpose of deriving the optimal conditions for the coupling strengths among individual components to maximize the absorption efficiency in the nanowires, we use temporal coupled-mode theory~(TCMT). 
The optimal conditions are exploited to identify suitable structural parameters. The functionality of actual devices is verified with full-wave simulation of PhC structures. We perform 2D simulations, with a geometry invariant in the third dimension, of four cascaded PhC cavities. Absorption efficiencies of about $80\%$ are achieved with a full width at half maximum (FWHM) of about $1$~nm. Three-dimensional~(3D) simulations of PhC slab structures are also carried out, showing similar spectral performance in the absorption. These 3D simulations are limited up to a cascade of two cavities due to the constrained computational memory. We expect that the proposed single-photon spectrometers using SNSPDs will be exploited in various on-chip applications where its practical usefulness is experimentally demonstrated in the near future.

\section{Proposed design of the spectrometer}
\begin{figure}[t]
\includegraphics[width=\linewidth]{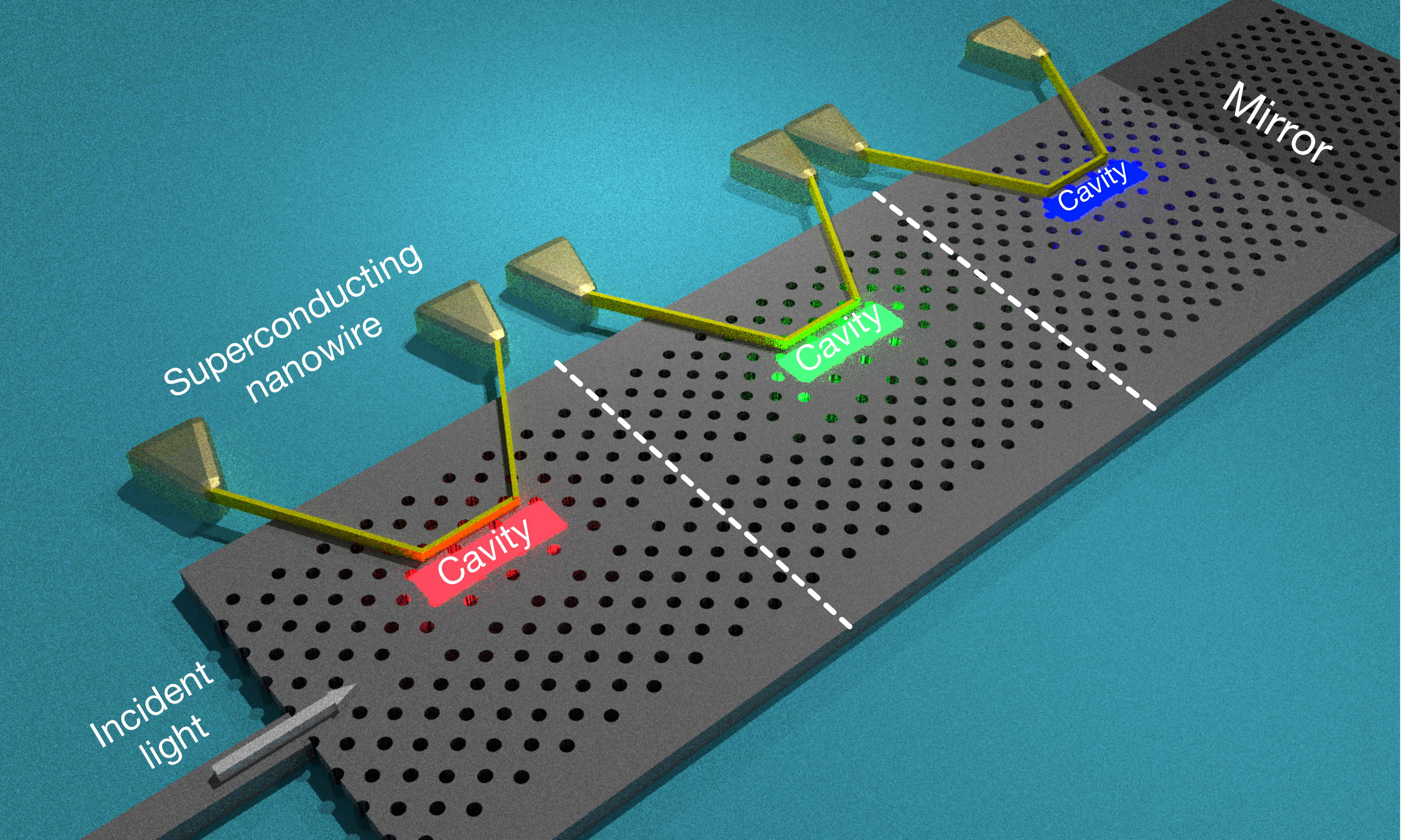}
\caption{Schematic of the proposed single-photon spectrometer using SNSPDs in cascaded PhC structures. 
The arrow indicates the injection of light to be spectrally measured, and the dashed lines represent the heterointerfaces between adjacent PhC structures with different lattice constants. A PhC mirror is placed at the end of the bus waveguide and reflection of the guided modes in a range of wavelengths of interest occurs. This back reflection is important to exceed efficiencies of 50\%. Individual point-defect PhC nanocavities are placed in each region next to the bus waveguide and superconducting nanowires are considered to be on top of the PhC slab and next to the cavities.}
\label{setup}
\end{figure}

The proposed design of our single-photon spectrometer using SNSPDs is depicted in Fig.~\ref{setup}. There, three PhC structures with three different lattice constants are exemplarily considered in a PhC slab. We design the PhC slab to be formed by air holes in a high-index dielectric material (e.g., Silicon) in view of experimental relevance.  The incident light is injected from an access waveguide to the bus waveguide. The injected light propagates through the bus waveguide without leaking to the PhC regions since the propagating modes shall exist inside the band gap~[see defect lines in Figs.~\ref{FigA1}(a) and (b)]~\cite{Mekis1996,Notomi2001}. The access waveguide can be designed in various ways to optimize the injection efficiency over a wide range of wavelength~\cite{Dutta2016}. Here we employ, for simplicity, a conventional strip waveguide. The PhC bus waveguide is side-coupled to point-defect PhC cavities with different resonance wavelengths, called L$x$-cavity with $x$ being the number of holes filled in a lateral direction~\cite{Akahane2003}. The evanescent coupling between the cavity and the bus waveguide occurs and, depending on the cavity resonance profile, enables the wavelength-selective demultiplexing scheme~\cite{Kuramochi2014a}. The cavity resonances are determined by its geometrical and material properties. Here, we tune the lattice constant of the PhC structures for a given material. Structures made from a sequence of such cavities are called heterostructures, whereas we call individual cavity structures a homostructure. Photons loaded in the cavity, transferred from the bus waveguide, can be absorbed by a superconducting nanowire that is placed on top of the PhC slab, next to the cavities. The photon absorption triggers detection events in the SNSPD. Such placement of the nanowires is intended to minimize the modification of the cavity resonance. The waveguide-cavity-nanowire coupled system constitutes a building block, leading to the final design by cascading those building blocks with different lattice constants that cause distinct resonance wavelengths of the cavities. The propagating waves are partially reflected and transmitted across the heterointerface (see dashed lines in Fig.~\ref{setup}) between adjacent PhC regions. Note that the PhC mirror placed at the end of the bus waveguide (see darker region in Fig.~\ref{setup}) is realized by a PhC structure with a lattice constant far detuned from the previous structures. Its presence allows to increase the absorption efficiency, in principle, up to 100\% in the absence of loss~\cite{Song2005a} by exploiting the concept of coherent perfect absorption. The role of the mirror will be much clearer in the next section when an analytical model of the structure is introduced.

\section{Analytical modeling}\label{Sec:TCMT}
The theoretical evaluation and prediction of the spectral performance requires an appropriate analytical model. Particularly, it shall enable us to find the conditions that allow us to identify the structural parameters of the single-photon spectrometer scheme that maximize the performance metric of the spectrometer. To be specific, we require a high efficiency and a fine spectral resolution, while the excellent timing characteristics comes from using extremely short superconducting nanowires. To express the conditions that lead to an optimal design, we employ here TCMT for a single waveguide-cavity-nanowire structure~\cite{Fan2003}. It will be shown that three such conditions exist.

\begin{figure}[b]
\includegraphics[width=0.8\linewidth]{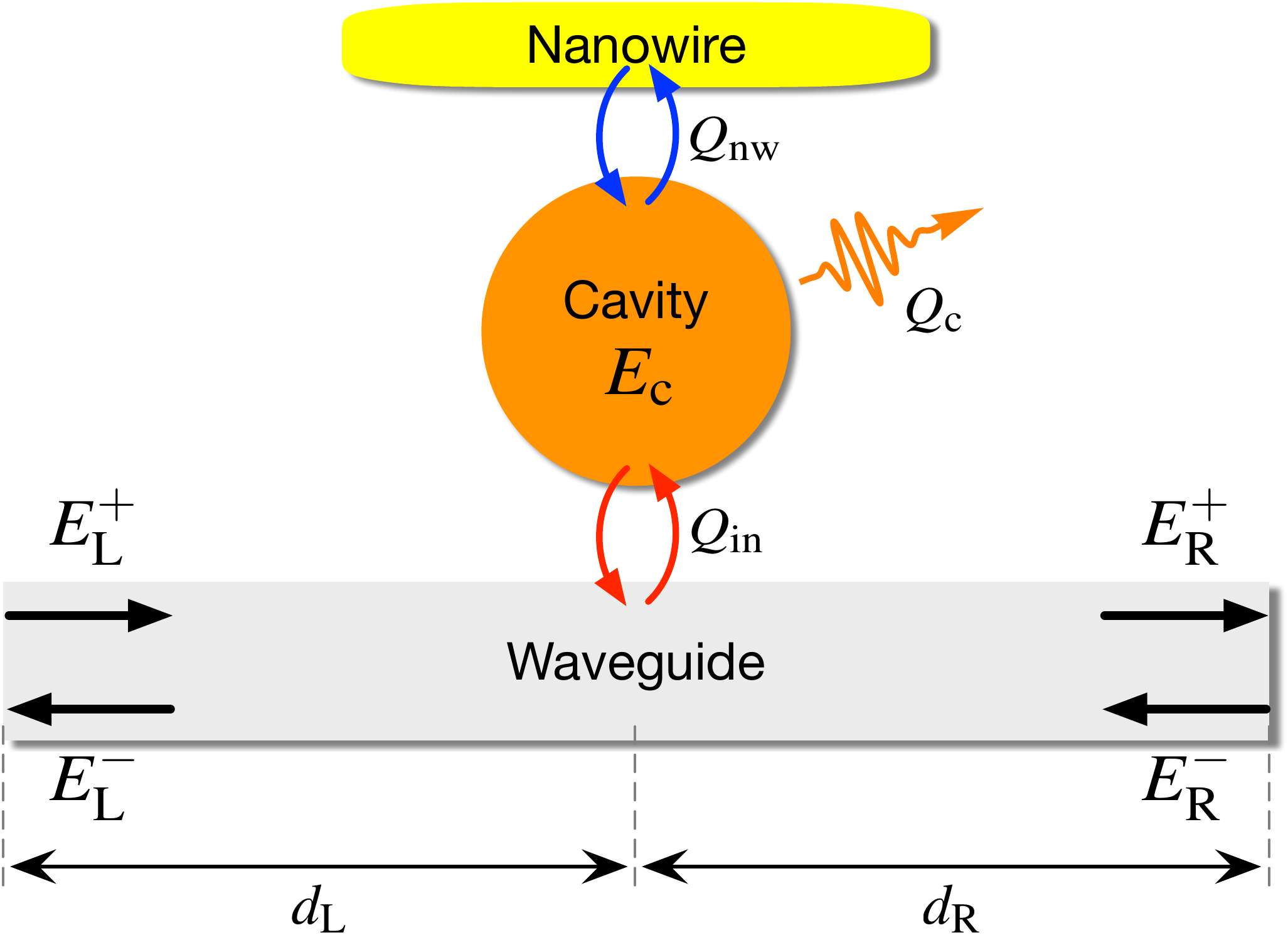}
\caption{A homostructure that consists of a waveguide, a cavity, and a nanowire. The figure illustrates also the notation used in the TCMT.}
\label{schematicHS}
\end{figure}

Consider a homostructure, where a cavity and a nanowire are placed in line and side-coupled to the bus waveguide in the middle, as illustrated in Fig.~\ref{schematicHS}. They are assumed to be coupled to each other via a point contact. $E_{\text{L}}^{+(-)}$ and $E_{\text{R}}^{+(-)}$ are the amplitudes of right-going (+) [left-going (-)] waves at the left (L) and right (R) side of the waveguide, respectively. Their relations can be written as
\begin{align}
E_{\text{L}}^{-}&=e^{-i\beta(d_\text{L}+d_\text{R})}E_{\text{R}}^{-}-\left( \frac{\omega_0}{2Q_\text{in}}\right)^{1/2}e^{-i\beta d_\text{L}}E_\text{c},\label{ELleft}\\
E_{\text{R}}^{+}&=e^{-i\beta(d_\text{L}+d_\text{R})}E_{\text{L}}^{+}-\left( \frac{\omega_0}{2Q_\text{in}}\right)^{1/2}e^{-i\beta d_\text{R}}E_\text{c},\label{ERright}
\end{align}
where $d_\text{L (R)}$ is the distance from the cavity to the left (right) boundary, $\beta$ is the frequency-dependent propagation constant of the mode in the bus waveguide, $\omega_0$ is the cavity resonance frequency, $Q_\text{in}$ is the quality factor associated with the coupling between the waveguide and the cavity, and $E_\text{c}$ is the amplitude of the cavity. The equation of motion for the amplitude $E_\text{c}$ reads as~\cite{Song2005a}
\begin{align}
\frac{dE_\text{c}}{dt}&=\left(i\omega_0-\frac{\omega_0}{2Q_\text{c}}-\frac{\omega_o}{2Q_\text{nw}}-\frac{\omega_0}{2Q_\text{in}} \right)E_\text{c} \nonumber\\
&\quad\quad+\left( \frac{\omega_0}{2Q_\text{in}}\right)^{1/2}e^{-i\beta d_\text{L}}E_{\text{L}}^{+} + \left( \frac{\omega_0}{2Q_\text{in}}\right)^{1/2}e^{-i\beta d_\text{R}}E_{\text{R}}^{-},
\label{cavityamplitude}
\end{align}
where the radiative loss, with an associated quality factor of $Q_\text{c}$, and the absorption in the nanowire, with an associated quality factor of $Q_\text{nw}$, are additionally taken into account. The radiated field ($E_\text{rad}$) and the field ($E_\text{nw}$) in the nanowire are written as
\begin{align}
E_\text{rad}&=\tilde{E}_\text{rad}-\left( \frac{\omega_0}{Q_\text{c}}\right)^{1/2}E_\text{c},\label{Erad}\\
E_\text{nw}&=\tilde{E}_\text{nw}-\left( \frac{\omega_0}{Q_\text{nw}}\right)^{1/2}E_\text{c},\label{Enw}
\end{align}
where the tilde denotes the fields at a time before the interaction with the cavity. They are assumed to be zero throughout
\footnote{The absence of a factor of $\sqrt{2}$ in the coupling terms can be explained when comparing it to the other coupling terms given in Eqs.~\eqref{ELleft} and \eqref{ERright}. Because the fields $E_\text{rad}$ and $E_\text{nw}$ are considered to be the superposed field of right- and left-propagating waves with the normalization factor $\sqrt{2}$, this cancels the other possible $\sqrt{2}$ factor in the coupling terms.}.

The set of above coupled equations [Eqs.~\eqref{ELleft} to \eqref{Enw}] describes the evolution of temporally coupled individual modes. It yields the relation between incoming and outgoing fields. To define a definite experimental situation, we consider the case where light impinges from the left, i.e., $E_{\text{R}}^{-}=0$. By eliminating the cavity field amplitude, one can define the transmission ($t$), reflection ($r$), loss ($l$), and absorption ($a$) coefficients~respectively as
\begin{align}
t=\frac{E_{\text{R}}^{+}}{E_{\text{L}}^{+}},~
r=\frac{E_{\text{L}}^{-}}{E_{\text{L}}^{+}},~
l=\frac{E_\text{rad}}{E_{\text{L}}^{+}},~
a=\frac{E_\text{nw}}{E_{\text{L}}^{+}},
\end{align}
for $E_\text{R}^{-}=0$.
Among those, the key quantity in this work is the absorption $\mathcal{A}=\vert a\vert^{2}$ in the nanowire, obtained by solving the above coupled equations. The absorption can be written as
\begin{align}
\mathcal{A}=\frac{\frac{1}{Q_\text{nw}}\frac{1}{2Q_\text{in}}}{\left|i\left(\frac{\omega}{\omega_0}-1\right)
	+\frac{1}{2Q_\text{c}}+\frac{1}{2Q_\text{nw}}+\frac{1}{2Q_\text{in}}\right|^2}.
\label{A}
\end{align}
One can see that the absorption $\mathcal{A}$ cannot be greater than $50\%$ since light extraction from the bus waveguide to the cavity is limited to maximally $50\%$. Such upper bound can also be explained by looking at the extent to which light is neither transmitted nor reflected, i.e., the total dissipation $\mathcal{D}=\mathcal{A}+\mathcal{L}$, where $\mathcal{L}=\vert l\vert^{2}$, written as
 \begin{align}
\mathcal{D} = \frac{\frac{1}{Q_\text{diss}}\frac{1}{2Q_\text{in}}}{\left\vert i\left(\frac{\omega}{\omega_{0}}-1\right)+ \frac{1}{2Q_\text{diss}}+\frac{1}{2Q_\text{in}}\right\vert^2},
\label{diss}
\end{align}
where $Q_\text{diss}$ is defined as
\begin{align}\label{Qdiss}
\frac{1}{Q_\text{diss}}&=\frac{1}{Q_\text{c}}+\frac{1}{Q_\text{nw}}.
\end{align}
Here, $\mathcal{D}\le 1/2$ and the upper bound is achieved when $Q_\text{in}=Q_\text{diss}$ at resonance. This is the first condition to be considered when optimizing individual homostructures in the spectrometer in the absence of a mirror at the end of the bus waveguide. This condition corresponds to the condition of critical coupling~\cite{Gorodetsky1999, Cai2000, Yariv2002}.
It can be easily verified that the remaining $50\%$ of light is equally distributed among transmission and reflection in the bus waveguide, i.e., $\mathcal{T}=\vert t\vert^{2}=1/4$ and $\mathcal{R}=\vert r\vert^{2}=1/4$.

Interestingly, the upper bound for $\mathcal{D}$ can be lifted up to unity by placing a mirror at the end of the bus waveguide, e.g., at the right edge in Fig.~\ref{schematicHS}. The presence of the mirror enables the complete destructive interference between the reflected light from the mirror and the reflected light from the cavity~\cite{Song2005a}. In other words, $\mathcal{T}=0$ and $\mathcal{R}=0$, so that all light can be transferred from the waveguide to the cavity-nanowire system. In the cavity-nanowire system, the light is then entirely dissipated. Such a scenario has been called coherent perfect absorption~\cite{Chong2010}.
The presence of the mirror imposes an additional relation of 
\begin{align}\label{mirror}
E_{\text{R}}^{-}&=e^{-i\Delta}E_{\text{R}}^{+},
\end{align}
where $\Delta$ is the phase delay introduced by the mirror. It can be shown that the modified dissipation $\mathcal{D}$ reads as
\begin{align}\label{D'}
\mathcal{D}
=\frac{\frac{1}{Q_\text{diss}Q_\text{in}^\text{eff}}}{\left\vert i\left(\frac{\omega-\omega_{0}^\text{eff}}{\omega_0}\right) +\frac{1}{2Q_\text{diss}}+\frac{1}{2Q_\text{in}^\text{eff}}\right\vert^2},
\end{align}
where the effective parameters modified by the mirror are given by 
\begin{align}\label{w0system}
\omega_0^\text{eff}&= \omega_0\left(1+\frac{\sin\theta}{2Q_\text{in}}\right), \qquad Q_\text{in}^\text{eff}=\frac{Q_\text{in}}{1+\cos\theta}.
\end{align}
Here, $\theta=2\beta d_\text{R} +\Delta$ is the phase difference between the left-propagating wave escaped from the cavity and the reflected wave from the mirror. It can be shown that $\mathcal{D}=1$ at resonance when $Q_\text{diss}=Q_\text{in}^\text{eff}$. This condition can be tuned by manipulating $\theta$, e.g., by controlling $\beta$, $d_\text{R}$, or $\Delta$. Note that $\beta$ and $\Delta$ depend on the wavelength. This is naturally accommodated in the full-wave simulation conducted in Sec.~\ref{FWS}. It is interesting to note that $\mathcal{D}\approx1$ can be kept in a broad range of~$\theta$. It implies that the extraction with a nearly unity efficiency is tolerant to moderate experimental imperfections~\cite{Song2005a}. It also alleviates the need for an extremely rigorous optimization for the distance $d_\text{R}$, i.e., roughly optimized parameters are acceptable for an approximately good performance.

When light is transferred to the cavity-nanowire system, the energy is distributed through radiation ($\mathcal{L}$) and absorption ($\mathcal{A}$). The extent to which light is absorbed in the nanowire, written as
\begin{align}
\mathcal{A}=\frac{\frac{1}{Q_\text{nw}Q_\text{in}^\text{eff}}}{\left\vert i\left(\frac{\omega-\omega_{0}^\text{eff}}{\omega_0}\right) +\frac{1}{2Q_\text{diss}}+\frac{1}{2Q_\text{in}^\text{eff}}\right\vert^2},
\label{Amirror}
\end{align}
increases with the ratio $Q_\text{c}/Q_\text{nw}$. This requires a high $Q$-factor for the cavity, i.e., $Q_\text{c}\gg Q_\text{nw}$, which is the second condition to be considered when optimizing the spectrometer over the structural parameters. 

The last important figure of merit for the spectrometer is its spectral resolution, i.e., a narrow FWHM is desired~\cite{Bogumil2003}. The absorption $\mathcal{A}$ of Eq.~\eqref{Amirror} yields the $\text{FWHM}_\omega$ in frequency written as
\begin{align}
{\rm FWHM}_\omega = \frac{\omega_{0}^\text{eff}}{Q_\text{total}},
\label{FWHMfreq}
\end{align}
where $Q_\text{total}$ is the total $Q$-factor given as 
\begin{align}\label{Qtotal}
\frac{1}{Q_\text{total}}=\frac{1}{Q_\text{diss}}+\frac{1}{Q_\text{in}^\text{eff}}.
\end{align}
Here $Q_\text{in}^\text{eff}$ is replaced by $Q_\text{in}$ in the absence of the mirror. 
The $\text{FWHM}_\omega$ of Eq.~\eqref{FWHMfreq} can also be written in wavelength as
\begin{align}
{\rm FWHM}_\lambda \approx \frac{\lambda_{0}^\text{eff}}{Q_\text{total}},
\end{align}
where $\lambda_{0}^\text{eff}$ is the effective resonance wavelength of the cavity and ${\rm FWHM}_\lambda\ll\lambda_{0}$ has been assumed. Obviously, high spectral resolution, represented by narrow $\text{FWHM}_{\lambda}$, requires $Q_\text{total}$ to be as large as possible, constituting the third condition to design the spectrometer.  

One can see that increasing $Q_\text{c}$ not only raises the absorption efficiency but also narrows the FWHM, i.e., a good cavity with high $Q_\text{c}$ is always favored. 
On the other hand, one can find a trade-off between the absorption efficiency and the FWHM with $Q_\text{nw}$, i.e., increasing $Q_\text{nw}$ decreases both the absorption efficiency and the FWHM.

Note that the spectral behavior of absorption $\mathcal{A}$ of Eqs.~\eqref{A}~and~\eqref{Amirror} follows a Lorentzian line shape, regardless of the presence of the mirror. This allows us to apply a Lorentzian fitting, without loss of generality, to the numerical data obtained by the full-wave simulation in the next section, to estimate the resonance wavelengths and FWHMs even when partial reflections occur across PhC heterointerfaces.~\footnote{As a remark, it is also worth to note that the formalism we use is valid only when the single-mode approximation holds or equivalently the temporal width of a propagating wave is extremely long with respect to the size of the whole structure. The formalism is thus not valid if the distance $d_\text{R}$ is considerably large, so that the behaviours of the multi-mode light interplay, making the phase $\theta$ frequency-dependent and subsequently the absorption spectra are no longer Lorentzian.}

\section{Full-wave simulations}\label{FWS}
The absorption in the proposed structure shown in Fig.~\ref{setup} is studied by numerical simulations of 2D geometries, comprising geometries invariant in the third dimension, and 3D slab geometries of actual structures feasible for fabrication. While 2D simulations allow the evaluation of an array of several PhC regions, 3D simulations are limited up to an array of two consecutive PhC regions due to the constrained computational memory. The actual design of a superconducting nanowire would be similar to the ones shown in Fig.~\ref{setup}, but we model it as a rectangular bar in both 2D and 3D geometries for simplicity while leaving investigation of complex actual designs to future work. In both simulations, we first tune the structural parameters to optimize the $Q$-factors according to the conditions just identified, and then perform the full-wave simulation to demonstrate the liability of our approach. All calculations presented here are done using the Waveoptics module of COMSOL MultiPhysics. The details of simulations and the results are explained below. 

\subsection{Two-dimensional photonic structures}
\begin{figure}[!t]
\includegraphics[width=1.05\linewidth]{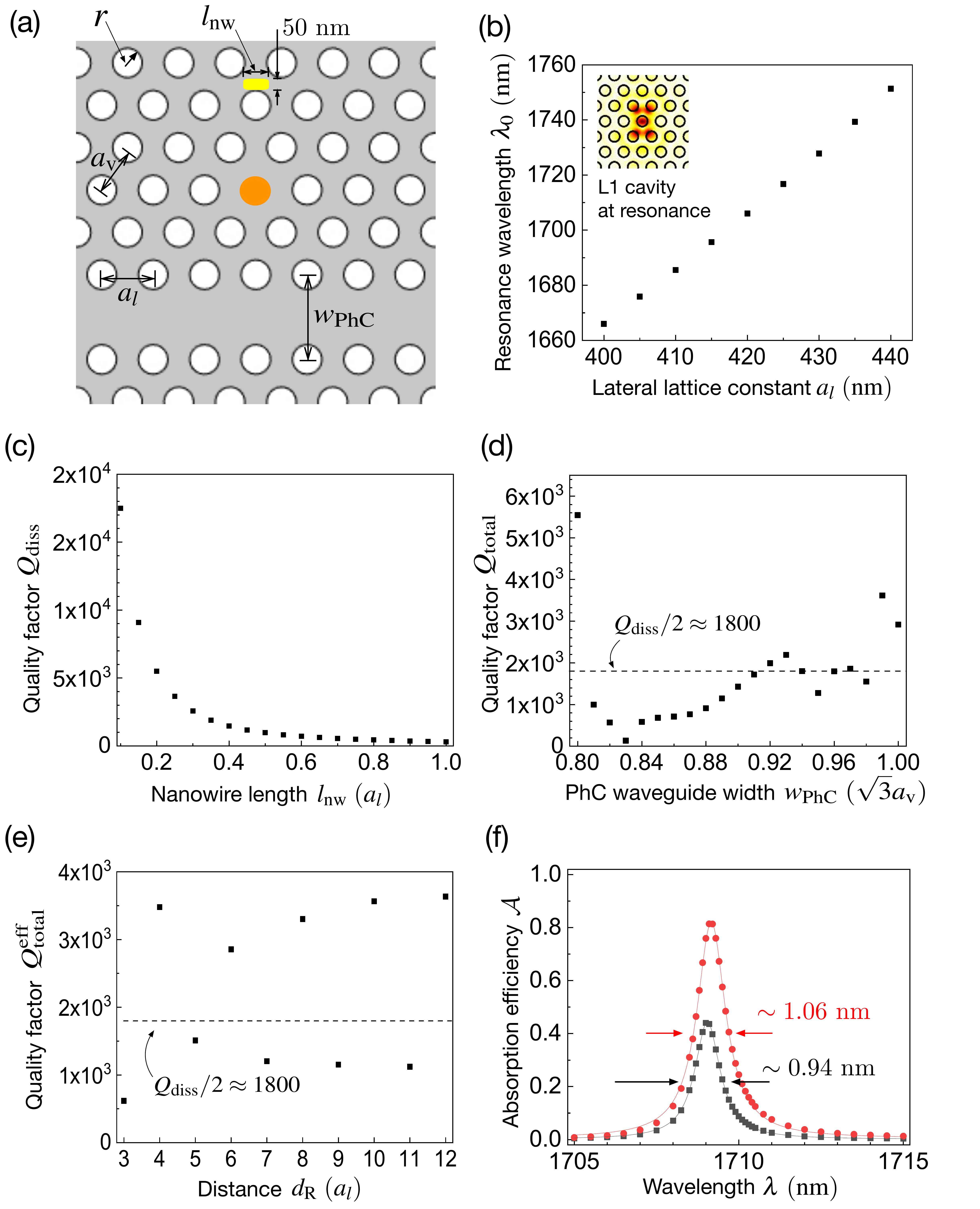}
\caption{(a)~Definitions of the geometrical parameters of the PhC homostructure. 
(b)~The resonance wavelength of an L$1$ cavity depending on the lateral lattice constant $a_\textit{l}$. Inset: the electric field amplitude profile of the fundamental L$1$ cavity mode. 
(c)~The $Q$-factor of the cavity-nanowire structure, $Q_\text{diss}$, depending on the length of the nanowire in units of $a_\textit{l}$ for $a_\textit{l}=420~\text{nm}$. 
(d)~The $Q$-factor of the waveguide-cavity-nanowire structure, $Q_\text{total}$ depending on the width of the PhC waveguide in units of $\sqrt{3}a_\textit{l}$, in comparison with $Q_\text{diss}$ obtained for $a_\textit{l}=420~\text{nm}$ and $l_\text{nw}=0.25a_\textit{l}$. 
(e)~The $Q$-factor of the waveguide-cavity-nanowire structure with a PhC mirror, $Q_\text{total}^\text{eff}$ depending on the distance $d_\text{R}$ between cavity center and the mirror interface in units of $a_\textit{l}$, in comparison with $Q_\text{diss}$ considered in panel (d). 
(f)~The spectra of absorption ${\cal A}$ in the nanowire in an optimized cavity in the absence (black squares) and presence (red circles) of the PhC mirror. Solid lines represent the Lorentzian fitting. An absorption efficiency of about $44\%$ at $1709.05~\text{nm}$ is obtained without the mirror, whereas $81\%$ is reached at the slightly modified wavelength of $1709.16~\text{nm}$ with the help of the mirror, where the constructive interference occurs through nearly satisfying the optimal conditions for $Q$-factors. The FWHMs are approximately $0.94~\text{nm}$ and $1.06~\text{nm}$, respectively. Here, $d_\text{R}=5a_\textit{l}=2.1~\mu\text{m}$ is chosen.}
\label{Fig3}
\end{figure}

%
Let us consider a 2D PhC structure with a triangular pattern of air holes, characterized by lateral and vertical lattice constants ($a_\textit{l}$ and $a_\text{v}$) and the radius of the hole ($r$), in Silicon as shown in Fig.~\ref{Fig3}(a). We set $a_\text{v}=400~\text{nm}$ and $r=120~\text{nm}$ throughout this work. The wavelength-dependent electric permittivity of Silicon is considered according to the Sellmeier equation with the parameters measured in Ref.~\cite{Salzberg1957}. A point-defect nanocavity with a missing hole (i.e., an L1 cavity) is considered and a NbN superconducting nanowire with a rectangular shape of size $l_\text{nw}\times50~\text{nm}$ is embedded next to the cavity [see yellow strip in Fig.~\ref{Fig3}(a)]. For the nanowire, the refractive index of $n_\text{NbN}=5.23+i5.82$, measured at $\lambda=1550~\text{nm}$~\cite{Hu2009}, is used for simulation~\footnote{The $n_\text{NbN}$ varies with the wavelength, e..g, $n_\text{NbN}=5.62+i6.14$ at $\lambda=1750~\text{nm}$~\cite{Hu2009}. For simplicity, a fixed value of $n_\text{NbN}$ measured at $\lambda=1550~\text{nm}$ is used in the whole range of wavelength considered in this work.}. The cavity-nanowire system is placed near the the PhC bus waveguide with a width of $w_\text{PhC}$.

The cavity resonance wavelength $\lambda_0$ can be tuned by changing the lateral lattice constant $a_\textit{l}$. The respective dependencies are shown in Fig.~\ref{Fig3}(b) for the L$1$ cavity. Such a feature enables to demultiplex distinct wavelengths carried in the bus waveguide by designing a cascade of structures of PhC with different lateral lattice constants and adding the respective cavities. Here we choose an L$1$ cavity since it has no higher-order resonance modes in the wavelength range of our interest. In Fig.~\ref{Fig3}(c), it is shown that the $Q$-factor of the cavity-nanowire system, $Q_\text{diss}$ of Eq.~\eqref{Qdiss}, decreases when increasing the length of the nanowire $l_\text{nw}$; shown here for a structure with $a_\textit{l}=420~\text{nm}$. This implies that the absorption within the nanowire increases with longer nanowires, as expected. From the latter, a length of $l_\text{nw}=0.25a_\textit{l}$, leading to $Q_\text{diss}\approx3650$, is chosen to match the $Q$-factor of the entire homostructure, $Q_\text{total}$, that can be tuned by varying the width $w_\text{PhC}$ of the PhC bus waveguide in units of $\sqrt{3}a_\text{v}$ [see Fig.~\ref{Fig3}(d)]. The first condition of optimal $Q$-factors read as $Q_{\rm diss}=Q_\text{in}$, which can be written as $Q_\text{total}=Q_\text{diss}/2$ via Eq.~\eqref{Qtotal}. This determines the width of the PhC bus waveguide as $w_\text{PhC}=0.91\sqrt{3}a_\text{v}$, for which $Q_\text{total}= 1720$. These considerations basically fix the geometrical details of the entire PhC structure. 

\begin{figure}[t]
	\includegraphics[width=1.1\linewidth]{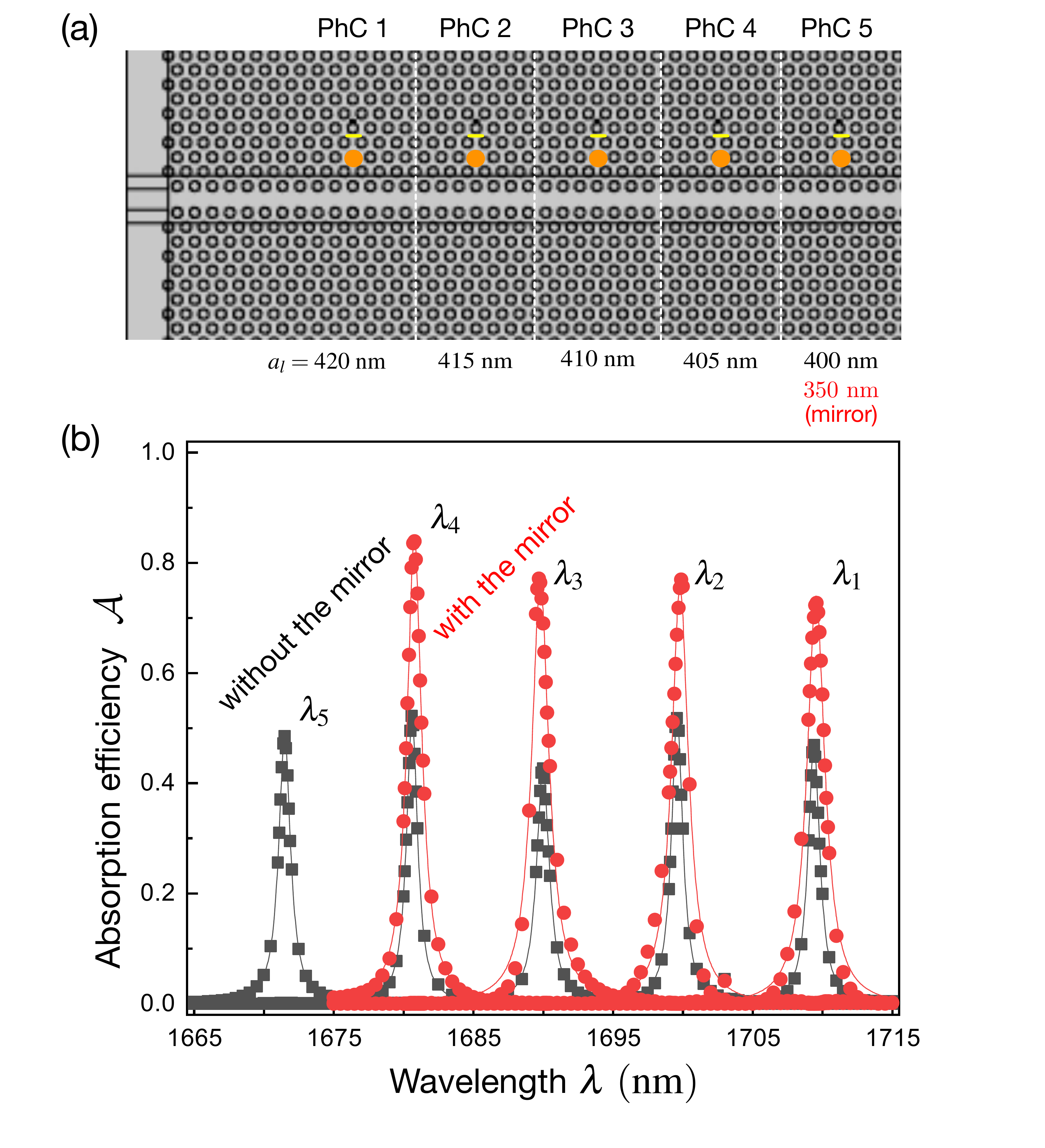}
	\caption{(a)~Geometrical details of the cascaded array of PhC-cavity-nanowire structures with the lateral lattice constant decreasing from the left to the right. For the case of the mirror being present, the lattice constant of the fifth structure (the rightmost part) is set to be far detuned from the last structure, prohibiting the mode matching of propagating modes between the PhC waveguides and thus causing full reflection. (b)~The absorption spectra ${\cal A}_j$ in the $j$th nanowire without (see black squares) and with the mirror turning the fifth structure into the far-detuned one (see red circles). Absorption efficiencies of $43\sim52\%$ are achieved in the absence of the mirror. This can be increased up to $74\sim84\%$ when the PhC mirror interface is implemented. Solid lines represent the Lorentzian fitting applied for the respective absorption peaks. The resonance wavelengths $\{\lambda_1,\lambda_2, \lambda_3, \lambda_4, \lambda_5\}$ in units of nm are found to be $\{1709.40, 1699.60, 1689.99, 1680.62, 1671.49\}$ without the mirror and $\{1709.54, 1699.80, 1689.80, 1680.78\}$ with the mirror, respectively. The FWHMs are about $1~\text{nm}$ at all absorption peaks. 
Clearly, the absorption in the different cavities can be seen that is key to the spectroscopic functionality. 
The unwanted absorption in the nearest neighboring nanowire amounts to only $0.1\sim0.2\%$, so that the crosstalk is negligible.}
\label{Fig4}
\end{figure}

For these chosen parameters, a full-wave simulation is performed, where the absorption ${\cal A}$ in the nanowire is calculated by integrating the power dissipated in the nanowire. It yields a maximal absorption ${\cal A}$ of about $44\%$ at $\lambda_0^\text{eff}=1709.05~\text{nm}$ with a FWHM of $0.94~\text{nm}$ [see black squares and curve in Fig.~\ref{Fig3}(f)]. The full spectral behaviors are shown in Fig.~\ref{FigA2}(a).

As analyzed in the previous section with the TCMT, the absorption efficiency can be further increased (in principle up to $100\%$) by placing a mirror at the end of the bus waveguide. Here, a PhC structure with $a_\textit{l}=350~\text{nm}$ is used as a PhC mirror at the end of the PhC bus waveguide. In such a structure, the total $Q$-factor is calculated while varying the distance $d_\text{R}$ between the center of the cavity and the heterointerface (or mirror interface) [see Fig.~\ref{Fig3}(e)], in order to find optimal parameters, for which the desired interference occurs, i.e., $Q_\text{total}^\text{eff}\approx Q_\text{diss}/2$. From this, $d_\text{R}=5a_\textit{l}$ is chosen, for which the absorption efficiency ${\cal A}$ increases up to about $81\%$ at $\lambda_0^\text{eff}=1709.16~\text{nm}$ while a FWHM of $1.06~\text{nm}$ is observed, as shown in Fig.~\ref{Fig3}(f) (see red circles and curve). One can see the full spectral behaviors in Fig.~\ref{FigA2}(b).

To simulate a multi-channel spectrometer, we consider a cascade of five homostructures with decreasing lateral lattice constants along the PhC bus waveguide ($a_\textit{l}=420, 415, 410, 405, 400~\text{nm}$), as shown in Fig.~\ref{Fig4}(a). It is shown that absorption efficiencies of $43\sim52\%$ with a FWHM of about $1~\text{nm}$ are achieved at five distinct resonance wavelengths [see black squares and curves in Fig.~\ref{Fig4}(b)]. Note that absorption efficiencies above $50\%$ are observed even in the absence of the mirror because of partial reflections that occur at the heterointerfaces between adjacent PhC structures. The absorption can be improved when the last PhC structure with $a_\textit{l}=400~\text{nm}$ is replaced by the far detuned one (i.e., $a_\textit{l}=350~\text{nm}$) in Fig.~\ref{Fig4}(a) for realizing the PhC mirror that causes full reflection. As a result, the absorption efficiencies are increased up to $74\sim84\%$, with a FWHM of $1~\text{nm}$, at the remaining four resonance wavelengths [see red circles and curves in Fig.~\ref{Fig4}(b)]. These results are mainly attributed to total reflection from the PhC mirror, but partial reflection also occurs at the heterointerfaces. The full spectral behaviors are presented in Figs.~\ref{FigA2}(c) and (d).

\subsection{Three-dimensional structures}

\begin{figure}[t]
\includegraphics[width=0.87\linewidth]{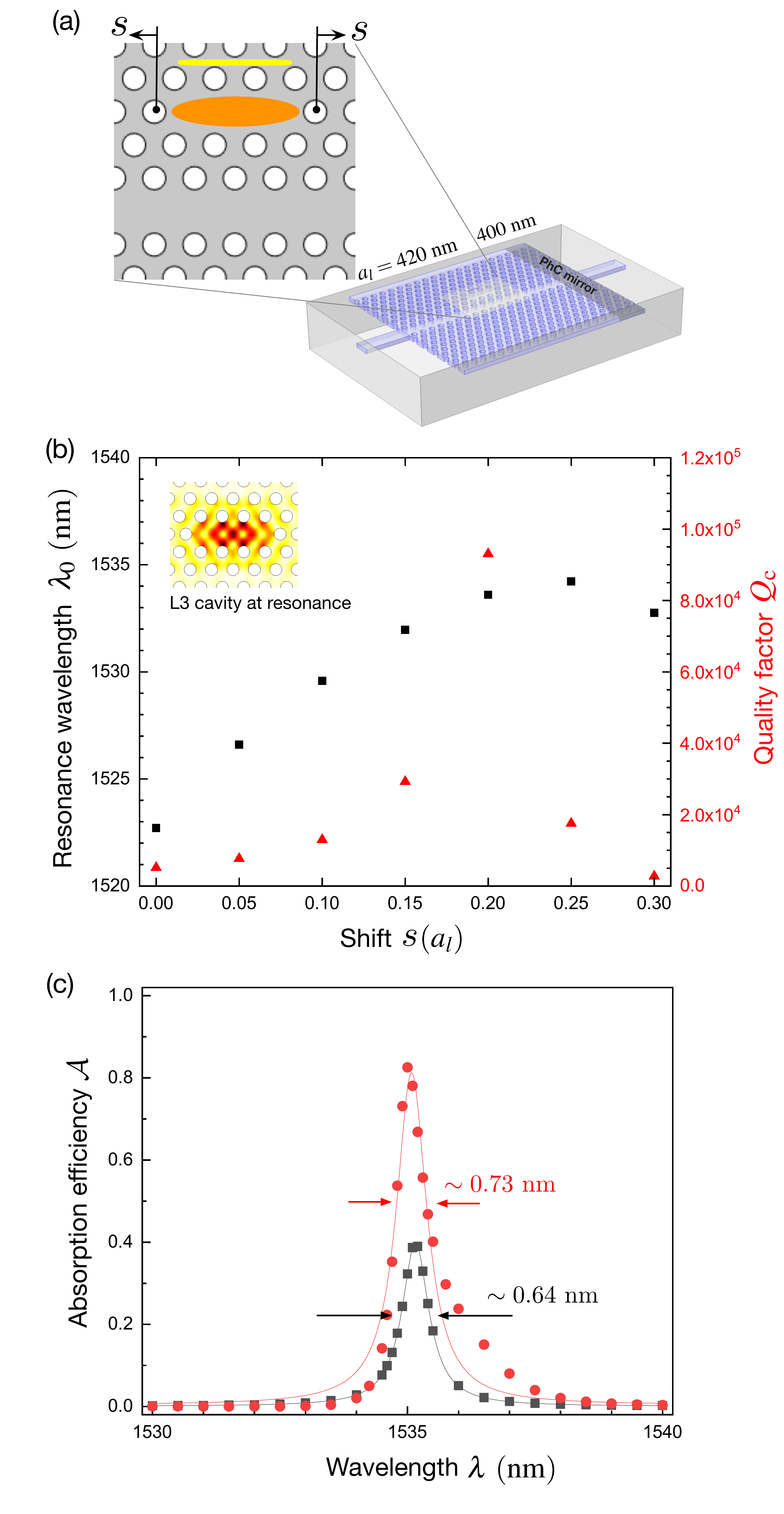}
\caption{
(a)~3D PhC slab structure with a height of $220~\text{nm}$, where an L$3$ cavity with the nearest neighbor holes displaced by $s$ is considered in PhC structures with $a_\textit{l}=420~\text{nm}$. The PhC mirror structure with $a_\textit{l}=400~\text{nm}$ is considered for the case exploiting the mirror.
(b)~The L$3$ cavity $Q$-factor and the resonance wavelength as a function of the displacement $s$ of the nearest neighbor holes. Inset: the electric field amplitude profile of the fundamental L3 cavity mode.
(c)~The absorption efficiencies as a function of wavelength with (black curve) and without (red curve) the PhC mirror for the structure that is moderately optimized (see Appendix~\ref{optimization} for the detail). Solid lines represent the Lorentzian fitting. An absorption efficiency of about 40\% at 1535.16~nm is obtained without the mirror, whereas about 80\% is reached at the slightly modified wavelength of 1535.08~nm with the help of the mirror. The FWHMs are approximately 0.64~nm and 0.73~nm, respectively. 
}
\label{Fig5}
\end{figure}

Contrary to the 2D PhC structures simulated above, PhC slab structures with a finite height are required to consider experimentally relevant scenarios. Here, the superconducting nanowire needs to be placed on top of the PhC slab and next to the cavity, as shown in Fig.~\ref{Fig5}(a). We assume in this work that the height of the PhC slab and the thickness of the nanowire are set to $220~\text{nm}$ and $4~\text{nm}$, respectively, similar to the structures fabricated in Ref.~\cite{Munzberg2018}. Above and below the PhC slab, we assume air. In the full-wave simulations, the entire structure is surrounded by perfectly matched layers that absorb outscatterred light. These geometrical modifications imply additional loss channels not only to the cavity but also to the PhC waveguide modes, modifying optical properties of the PhC structures.

With these additional loss channels, the $Q$-factor of a point-defect L$x$ PhC nanocavity is significantly reduced (e.g, $Q_\text{c}\approx 3980$ for the L$1$ cavity) when compared to 2D structures made by the same defect number $x$. The $Q$-factor increases generally with $x$, and we choose here a L$3$ cavity since cavities with $x>3$ exhibit multiple resonances in the relevant wavelength range. We wish to avoid this to suppress any spectral crosstalk that would harm the ability of our spectrometer.
 
The L$3$ cavity $Q$-factor can be further increased by adjusting the position of the neighboring holes~\cite{Akahane2005}, but it requires significant efforts in the optimization to reach the maximum $Q$-factor for a given $x$~\cite{Kuramochi2014b}. The optimization over several neighboring holes is an option, but in this work we only consider the shift $s$ of the nearest neighbor holes [see Fig.~\ref{Fig5}(a)]. This has been shown to rely on the most dominant factor to modify the $Q$-factor by orders of magnitude~\cite{Akahane2003} and is sufficient for the proof-of-principle numerical demonstration of this work. The shift of the nearest neighbor holes modifies not only the $Q$-factor of the L$3$ cavity but also its resonance wavelength, as shown in Fig.~\ref{Fig5}(b). A $Q$-factor of about $9\times10^4$ is obtained at the resonance wavelength of $1533.6~\text{nm}$ for a shift of $s=0.2a_\textit{l}$, where $a_\textit{l}=420~\text{nm}$ is chosen. The same procedure considered in Figs.~\ref{Fig3}(c) to (e) is performed to optimize the geometrical parameters in 3D structures (see Appendix~\ref{optimization}), resulting in $l_\text{nw}=2a_\textit{l}$ yielding $Q_\text{diss}\approx~3390$ and $w_\text{PhC}=0.66\sqrt{3}a_\textit{l}$ leading to $Q_\text{total}\approx1390$ in the absence of the mirror. For structures with the PhC mirror, an additional PhC structure with $a_\textit{l}=400~\text{nm}$, $d_\text{R}=10a_\textit{l}$ is chosen, for which $Q_\text{total}^\text{eff}\approx1450$ is obtained. For these chosen parameters, it is shown that the maximum absorption efficiency ${\cal A}$ of about $80\%$ ($40\%$) is achieved in the presence (absence) of the mirror at the resonance wavelength of $1535.08~\text{nm}$ ($1535.16~\text{nm}$) with a FWHM of $0.73~\text{nm}$ ($0.64~\text{nm}$) [see red circles (black squares) and curves in Fig.~\ref{Fig5}(c)]. Note that the measured wavelength range is different from that observed in 2D simulation since larger cavities are considered in 3D simulation. This result demonstrates that the proposed cavity-nanowire-waveguide 3D structure can resolve wavelengths of the light when appropriate optimal design conditions are satisfied, as expected from analytical modeling introduced in Sec.~\ref{Sec:TCMT}. Slightly lowered absorption efficiencies in comparison with the results shown in Fig.~\ref{Fig3}(f) are attributed to the presence of additional losses into free space, eventually increasing the loss~${\cal L}$. The full spectral behaviors are shown in Figs.~\ref{FigA2}(e) and (f).
\begin{figure}[t]
\includegraphics[width=1.1\linewidth]{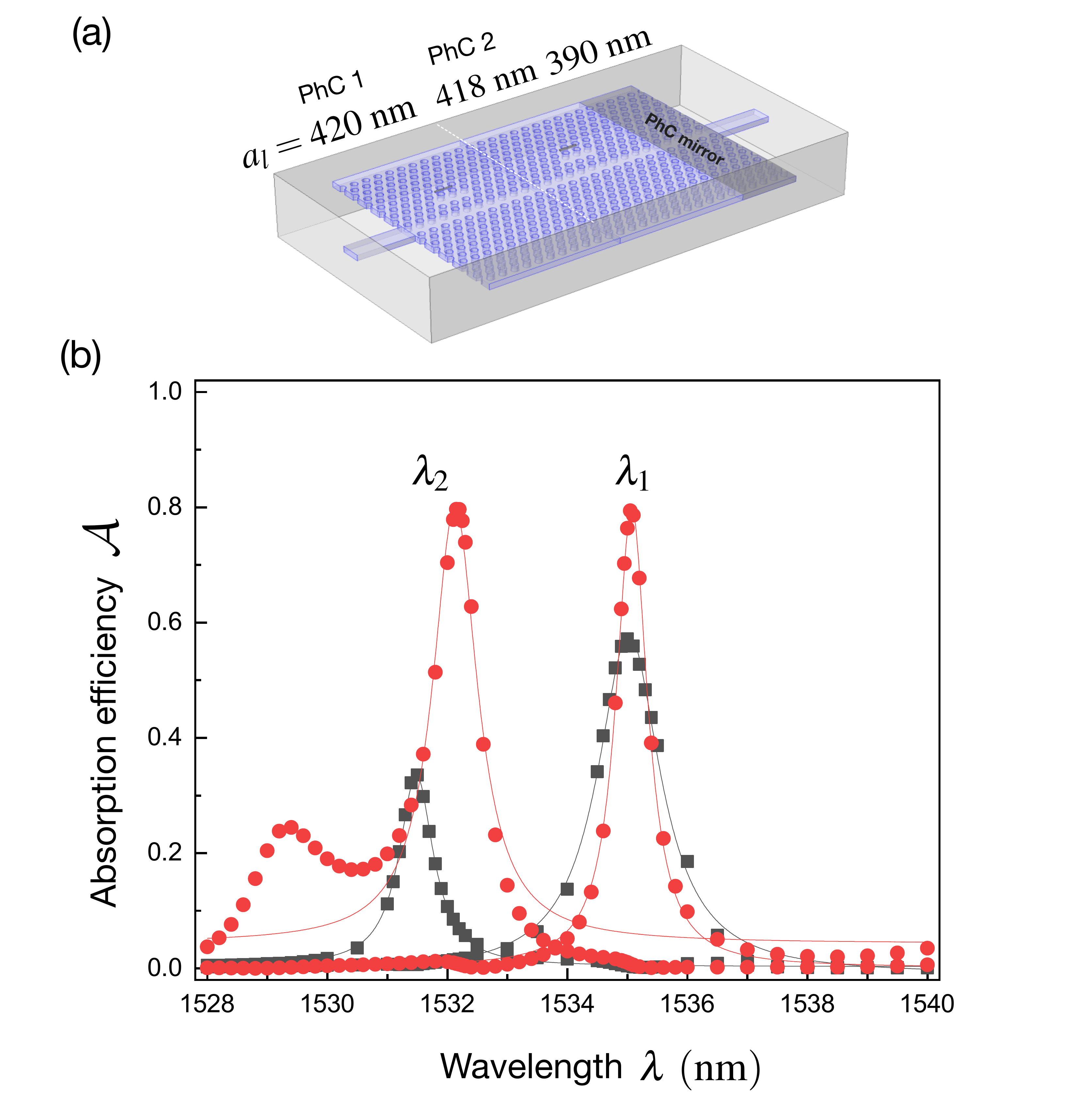}
\caption{
(a)~3D PhC slab structure with $220~\text{nm}$ height, where two L$3$ cavities in PhC structures with $a_\textit{l}=420, 418~\text{nm}$ and the PhC mirror structure with $a_\textit{l}=390~\text{nm}$ are considered. 
(b)~
The spectral dependent absorption ${\cal A}_j$ in the $j$th nanowire in the absence (see black squares) and presence (see red circles) of the PhC mirror. Solid lines represent the Lorentzian fitting applied to the respective absorption peaks.
The resonance wavelengths $\{\lambda_1,\lambda_2\}$ in units of nm are found to be $\{1535.03, 1531.48\}$ without the mirror and $\{1535.07, 1532.12\}$ with the mirror, respectively. The FWHMs are about $1~\text{nm}$ at all the absorption peaks. 
The unwanted absorption in the nearest neighboring nanowire amounts to only $0.4\sim0.8\%$, slightly larger than the case of 2D structures, but still negligibly small. Note that a small red peak is observed around $\lambda\approx1529~\text{nm}$ due to a Fabry-Perot effect occurring between the input facet and the PhC mirror interface~\cite{Song2008}. This effect could be alleviated by optimizing the structure of the access waveguide to reduce the reflection from input facet~\cite{Miyai2004, Vlasv2005}.
}
\label{Fig6}
\end{figure}

For multi-channel 3D simulations, we are limited to a small number of channels since simulation of many channels (e.g. more than two) is computationally prohibitive due to memory requirements. We thus perform 3D simulations for two cascaded structures without the PhC mirror, and with the PhC mirror region additionally attached to the end of the structure [see Fig.~\ref{Fig6}(a)]. Here, the lateral lattice constants are set to $420~\text{nm}$ and $418~\text{nm}$ for individual PhC structures, and $390~\text{nm}$ for the PhC mirror structure. Note that here the mirror structure with $a_\textit{l}=390~\text{nm}$, larger than $a_\textit{l}=350~\text{nm}$ used for 2D structures, is chosen due to the narrowed range of the line-defect guided modes [see Fig.~\ref{FigA1}(b)]. In Fig.~\ref{Fig6}(b), we show that absorption efficiencies ${\cal A}$ of about $80\%$ ($35\sim60\%$) are obtained with FWHMs of $0.64~\text{nm}$ ($1.29~\text{nm}$) and $0.99~\text{nm}$ ($0.69~\text{nm}$) at the resonance wavelengths of $\lambda_1=1535.07~\text{nm}$ $(1535.03~\text{nm})$ and $\lambda_2=1532.12~\text{nm}$ $(1531.48~\text{nm})$ in the presence (absence) of the mirror, respectively. 
The effect of partial reflections occurring at the heterointerfaces is incorporated in the results, which has led the absorption efficiency to be slightly greater than 50\% even when the PhC mirror is absent.
The full spectral behaviors are shown in Figs.~\ref{FigA2}(g) and (h).

\section{conclusion}
We have proposed a design of a superconducting nanowire single-photon spectrometer that exploits cascaded photonic crystal cavities. They are connected by a photonic crystal bus waveguide that delivers the light to the respective cavities. On the base of an analytical model that relies on TCMT, we have discussed the optimal conditions required to maximize the absorption efficiencies in the nanowires placed next to the PhC nanocavities, and showed that these optimal conditions are attainable by means of full-wave simulations for both 2D and 3D designs. It has thus been shown that absorption efficiencies of about $80\%$ in the nanowires can be achieved in full-wave simulations despite the presence of the scattering loss that occurs from injection, and radiation losses from the cavity and the PhC waveguide to free space. 

The analytical and numerical evidences we have provided through this work will motivate further investigations in the near future. 
For example, one can study similar PhC structures which keep the same lattice constant for all sub-regions, where the resonance wavelength of the PhC cavity is tuned differently by modifying the positions or radii of the holes near the cavity region~\cite{Gong2010}.
The experimental implementation of the proposed design relying on 3D air-bridged PhC slab structures requires the development of novel experimental methods to avoid potential damages to superconducting nanowires, which would occur when applying canonical fabrication techniques such as hydrofluoric acid wet etching. Employing the established methods that the PhC slab structures and superconducting nanowires are fabricated on top of a substrate~(e.g., $\text{SiO}_2$ used in a relevant experiment~\cite{Munzberg2018}), is possible with current technology. One should, however, take care of $Q$-factors to be modified by the presence of a substrate. For example, the $Q$-factor of the point-defect PhC cavity on a substrate is significantly degraded~\footnote{We have calculated the $Q$-factors of L1~(with the shift of s=0.2), L3~(s=0.2), and L5~(s=0.25) cavities on $\text{SiO}_2$ substrate and they are 771, 983, and 2356, respectively. These values are much smaller than the $Q$-factors of L1~(s=0.2), L3~(s=0.2), and L5~(s=0.25) cavities on air substrate, which are 3975, 93046, and 114856, respectively.}. On substrates, $Q$-factor can be enhanced via fine tuning of the position of the neighboring holes~\cite{Akahane2005, Kuramochi2014b}. 
It is also important to consider additional structural optimization for the access waveguide to achieve highly and uniformly efficient injection of a broadband light to the whole device structure~\cite{Dutta2016}.

We expect these improvements in experiment and theory to facilitate the use of the proposed single-photon spectrometers for a variety of quantum information processing applications in photonic crystal platforms with diverse functionalities which have already been developed over the last few decades.

\section*{Acknowledgments}
Y.~Yun thanks Aimi Abass and Julian M{\" u}nzberg for helpful comments in numerical simulation. We gratefully acknowledge partial financial support by the Deutsche Forschungsgemeinschaft (DFG, German Research Foundation) Project-ID 258734477 SFB 1173, from the European Union's Horizon 2020 research and innovation program under the Marie Sk\l odowska-Curie grant agreement \text{No. 675745}, and by the VolkswagenStiftung. A.~Vetter acknowledges support by the Karlsruhe School of Optics and Photonics (KSOP).

\appendix
\section*{Appendix}

\section{Photonic band structures}\label{AppendixA}
\setcounter{equation}{0}
\renewcommand{\theequation}{A\arabic{equation}}
\setcounter{figure}{0}
\renewcommand{\thefigure}{A\arabic{figure}}
By means of plane-wave expansion method available in MIT Photonic-Bands (MPB), we obtain not only the PhC band diagrams of 2D and 3D PhC slab structures, but also the dispersion curves of the line-defect guided modes for the PhC waveguide. Figures~\ref{FigA1}(a) and (b) show the photonic modes in the 2D and 3D slab structures with $a_\textit{l}= 420~\text{nm}$, $a_\text{v}=400~\text{nm}$, and $r=120~\text{nm}$. The photonic band gaps are found in a range of wavelength between $1526.3~\text{nm}$ and $1985.9~\text{nm}$ for 2D structures and between $1265.7~\text{nm}$ and $1585.6~\text{nm}$ for 3D structures [see yellow regions], inside which the guided even- and odd-modes lie [see red and blue dots, respectively]. Note that in 3D slab structures only the odd-mode mode exists in the range of our interest. 
\begin{figure}[t]
	\includegraphics[width=1.1\linewidth]{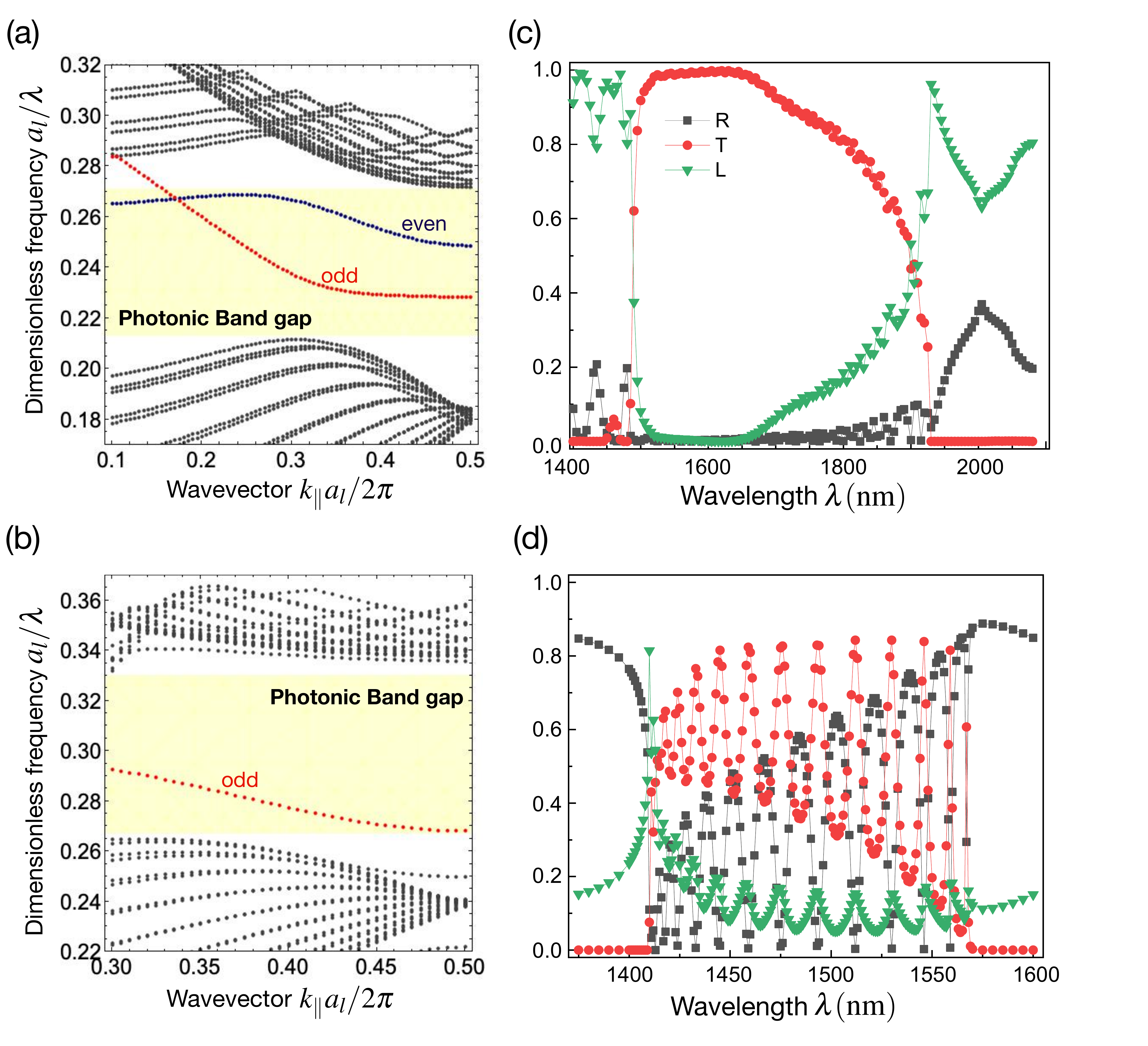}
	\caption{
Photonic band structures for a 2D structure in~(a) and a 3D structure in~(b) as a function of in-plane wavevector.
Panels (c) and (d) show the transmission, reflection, and loss through the 2D structure considered in panel~(a) and the 3D structure considered in panel~(b), respectively. Here the light is injected through the access waveguide as shown in Fig.~\ref{setup}.
Lines connecting dots are to guide the eyes.
}
	\label{FigA1}
\end{figure}
\begin{figure*}[]
\includegraphics[width=1\linewidth]{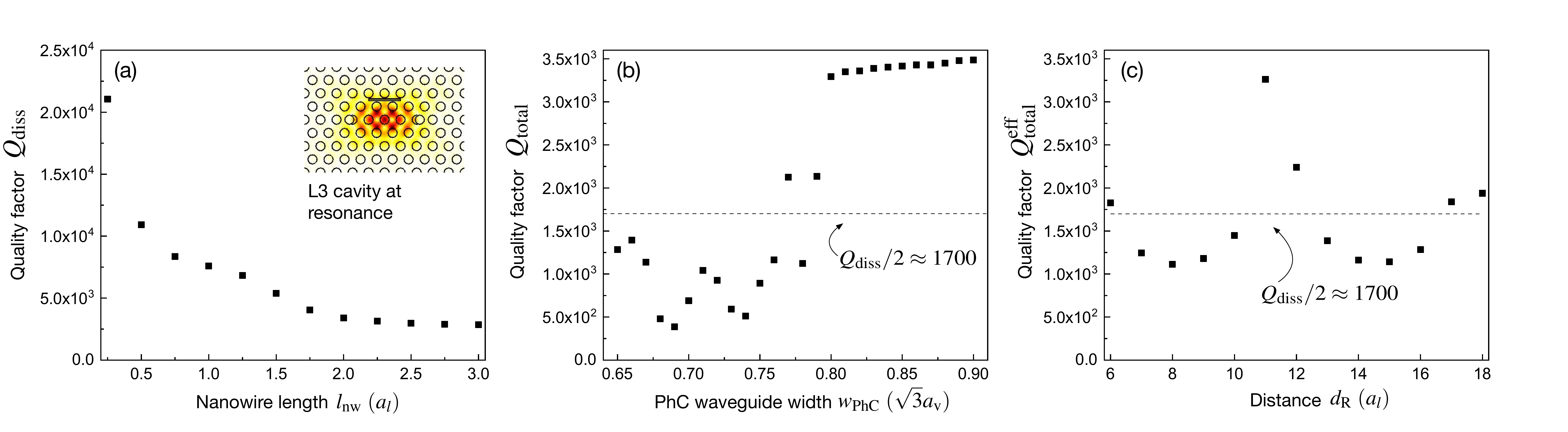}
\caption{
(a)~The $Q$-factor of the cavity-nanowire structure, $Q_\text{diss}$, with the length of the nanowire in units of $a_\textit{l}$ for $a_\textit{l}=420~\text{nm}$. The inset presents the electric field amplitude profile in the L3 cavity at resonance.
(b)~The $Q$-factor of the waveguide-cavity-nanowire structure, $Q_\text{total}$, with a width of the PhC waveguide in units of $\sqrt{3}a_\textit{l}$, in comparison with $Q_\text{diss}$ obtained for $a_\textit{l}=420~\text{nm}$ and $l_\text{nw}=2a_\textit{l}$. 
(c)~The $Q$-factor of the waveguide-cavity-nanowire structure in the absence of the PhC mirror, $Q_\text{total}^\text{eff}$, with a distance between the cavity center and the mirror interface in units of $a_\textit{l}$, in comparison with $Q_\text{diss}$ considered in panel (b). 
}
\label{FigA3}
\end{figure*}

\begin{figure*}[]
\includegraphics[width=1\linewidth]{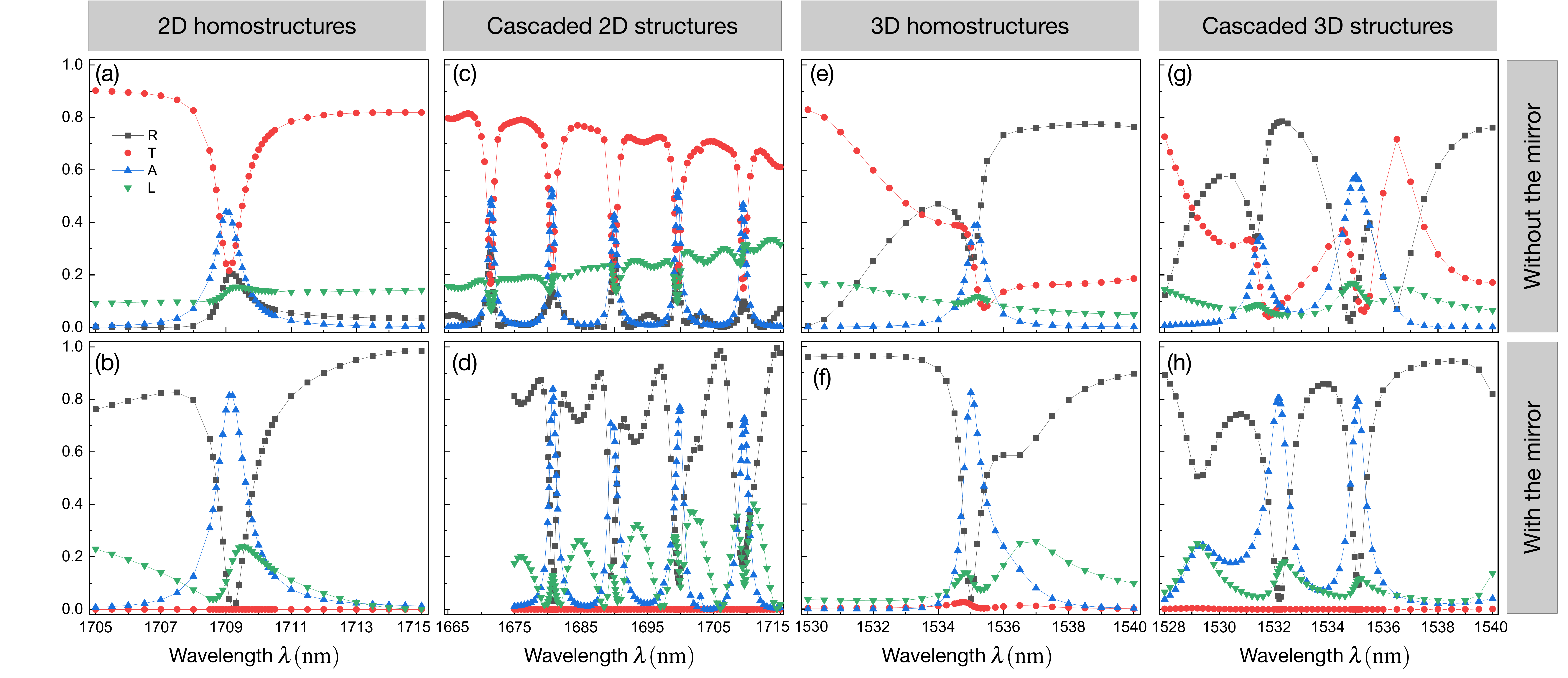}
\caption{The full spectral behaviors of reflection, transmission, absorption, and loss in 2D homostructures considered in Fig.~\ref{Fig3}(f), cascaded 2D structures considered in Fig.~\ref{Fig4}(b), 3D homostructures considered in Fig.~\ref{Fig5}(c), and cascaded 3D structures considered in Fig.~\ref{Fig6}(b) are calculated via the full-wave simulation in the cases that the PhC mirror is absent (see upper panel) or present (see lower panel). Lines connecting dots are to guide the eyes.}
	\label{FigA2} 
\end{figure*}

In the full-wave simulation in Sec.~\ref{FWS}, we exploit only the fundamental TE (or odd) mode since transmission of the TM (or even) mode in the PhC waveguide is strongly inhibited due to the large group index that inevitably increases losses and distortion of the pulse~\cite{Baba2008,Joannopoulosbook}. The full-wave simulation is also performed for the PhC waveguide structures in the absence of cavity and nanowire in order to understand the background spectral behaviors. Figures~\ref{FigA1}(c) and (d) present the background spectral behaviors of the light that is injected through the access waveguide. Note that the effect of a Fabry-Perot resonator formed between the input interface and the mirror interface induces the splitting of the absorption peaks~\cite{Song2008}. This explains the small red peak observed at around $\lambda\approx 1529~\text{nm}$ in Fig.~\ref{Fig6}(b) in 3D structures. No splitting is observed in 2D structures [see Fig.~\ref{Fig4}(b)] since the effect of Fabry-Perot is less significant as shown in Fig.~\ref{FigA1}(c). The reflection, modulating with wavelength, occurred by the Fabry-Perot effect is unwanted since it limits the maximum absorption efficiency in the nanowire. 
The Fabry-Perot effect can be significantly minimized by optimizing the structure of the access waveguide, consequently resulting in highly and uniformly efficient injection of a broadband light to the whole device structure~\cite{Dutta2016}.

\section{Parameter turning for 3D PhC slab structures}\label{optimization}
Similarly to the exploration made in 2D structures [see Figs.~\ref{Fig3}(c)-(e)], we also investigate the $Q$-factors with varying various geometric parameters in 3D structures, consequently leading to the results shown in Fig.~\ref{Fig6}(b). First, we examine the dependence of the length of the nanowire in the $Q$-factor of the waveguide-cavity structure, $Q_{\rm diss}$, for the L$3$ cavity with $s=0.2a_\textit{l}$ chosen from Fig.~\ref{Fig5}(b). The result is shown in Fig.~\ref{FigA3}(a), which leads us to choosing $l_\text{nw}=2a_\textit{l}$, yielding $Q_{\rm diss}\approx3390$. Secondly, the calculation of the total $Q$-factor $Q_\text{total}$ is performed with changing the width of the PhC waveguide. It is shown in Fig.~\ref{FigA3}(b), from which we choose $w_{\rm PhC}=0.66\sqrt{3}a_{\rm v}$ and it yields $Q_{\rm total}\approx 1390$. Finally, we calculated the $Q$-factor $Q_\text{total}^\text{eff}$ modified by the presence of the PhC mirror with $a_\textit{l}=400~\text{nm}$ with varying the distance $d_\text{R}$ between the cavity center and the mirror interface. The result is shown in Fig.~\ref{FigA3}(c), from which $d_\text{R}=10a_\textit{l}$ is chosen as it leads to $Q_{\rm total}^{\rm eff}\approx 1450$. 

The full spectral behaviors of reflection, transmission, absorption, and loss in 2D homostructures considered in Fig.~\ref{Fig3}(f), cascaded 2D structures considered in Fig.~\ref{Fig4}(b), 3D homostructures considered in Fig.~\ref{Fig5}(c), and cascaded 3D structures considered in Fig.~\ref{Fig6}(b) are shown in Fig.~\ref{FigA2} when the PhC mirror is absent (see upper panel) or present (see lower panel).

\newpage
\newpage
\bibliography{reference}

\end{document}